\documentclass[12pt,preprint]{aastex}
\usepackage{emulateapj5}
\usepackage{onecolfloat}
\usepackage{graphicx}
\usepackage{times,psfig}

\newcommand{\sm}{\, {\rm M}_{\odot}}
\newcommand{\sL}{\, {\rm L}_{\odot}}
\newcommand{\kms}{\, {\rm km~s}$^{-1}$\,}
\def\gsim { \lower .75ex \hbox{$\sim$} \llap{\raise .27ex \hbox{$>$}} }
\def\lsim { \lower .75ex \hbox{$\sim$} \llap{\raise .27ex \hbox{$<$}} }

\shorttitle{Tidal debris in the Galactic disk}
\shortauthors{A. Helmi et al.}

\begin{document}

\twocolumn[

\title{On the nature of the ring-like structure in the outer Galactic disk}

\author{Amina Helmi\altaffilmark{1},}
\affil{Astronomical Institute Utrecht, P.O.Box 80000, 3508 TA Utrecht,
The Netherlands}

\author{Julio F. Navarro\altaffilmark{2}, Andr\'es Meza\altaffilmark{3}}
\affil{Department of Physics and Astronomy, University of Victoria, 
Victoria, BC V8P 1A1, Canada}

\author{Matthias Steinmetz\altaffilmark{4}} 
\affil{Astrophysikalisches Institut Potsdam, An der Sternwarte 16,
D-14482 Potsdam, Germany, and Steward Observatory, 933 North Cherry
Avenue, Tucson, AZ 85721, USA}


\author{Vincent R. Eke\altaffilmark{5}}
\affil{Physics Department, Durham University, South Road, Durham DH1 3LE, England}

\begin{abstract}
We examine the tidal disruption of satellite galaxies in a
cosmological simulation of the formation of a disk galaxy in the
$\Lambda$CDM scenario. We find that the disruption of satellite
galaxies in orbits roughly coplanar with the disk leads naturally to
the formation of ring-like stellar structures similar to that recently
discovered in the outer disk of the Milky Way.  Two interpretations
appear plausible in this context. One is that the ring is a
transitory, localized radial density enhancement reflecting the
apocenter of particles stripped from a satellite during a recent
pericentric passage (a ``tidal arc'' reminiscent of the tidal arms
seen in disk galaxy mergers).  In the second scenario, the ring is
analogous to the ``shells'' found around some elliptical galaxies, and
would result from a minor merger that took place several Gyr ago. The
two interpretations differ in several ways. Tidal arcs are expected to
span a limited longitude range; may carry a substantial fraction of
the original mass of the satellite; should exhibit a significant
velocity gradient with Galactic longitude; and are in all likelihood
asymmetric above and below the plane of the disk. Shells, on the other
hand, may comprise at most a small fraction of the original mass of
the satellite and, due to their more relaxed state, ought to be
symmetric above and below the plane, with no discernible velocity
gradients across the structure.   If confirmed as a tidal feature,
the ring discovered by SDSS in the outer Galactic disk would
strengthen the view---supported by numerical simulations---that minor
mergers have played a significant role building
up not only the stellar halo, but also the disk components of the
Galaxy.
\end{abstract}

]
\altaffiltext{1}{NOVA Fellow; a.helmi@phys.uu.nl}
\altaffiltext{2}{Fellow of CIAR and of the Alfred P. Sloan Foundation; jfn@uvic.ca}
\altaffiltext{3}{ameza@uvic.ca}
\altaffiltext{4}{David and Lucile Packard Fellow and Sloan Fellow; msteinmetz@aip,de}
\altaffiltext{5}{Royal Society University Research Fellow; V.R.Eke@durham.ac.uk}

\section{Introduction}
\label{sec:intro}

Large-scale surveys like the Sloan Digital Sky Survey (SDSS) are
starting to probe the limits of our understanding of the structure and
dynamics of the Galaxy (Newberg et al. 2002).  The most recent example
is the discovery of a coherent structure at large
Galactocentric distance and low galactic latitude spanning about 100
degrees on the sky (Yanny et al. 2003, hereafter Y03; Ibata et
al. 2003, hereafter I03). The nature of this ``ring'' of low
metallicity stars remains a puzzle.  Y03 suggest that it could be the
debris of a tidally disrupted satellite moving on roughly circular
orbits, while I03 prefer the hypothesis that the ring results from
perturbations in the disk such as the remnants of ancient
warps. Neither scenario is clearly preferable. Could the debris of a
disrupted galaxy remain a sharp feature at 20 kpc from the Galactic
center while being stretched over one hundred degrees on the sky?
What kind of disk instability would put metal-poor stars on such a
distant ring around the Galaxy?

It is possible to shed some light on the viability of such scenarios
by turning to cosmological simulations of galaxy formation. Over the
past few years, these simulations have improved in resolution to the
point of allowing a detailed analysis of morphological and structural
features of galaxies and their relation to individual merger and
accretion histories (Bekki \& Chiba 2001; Springel \& Hernquist 2003;
Governato et al. 2002; Sommer-Larsen et al. 2002; Abadi et
al. 2002a,b). Abadi et al. (2002b), in particular, point out that the
bulk of old disk stars may originate from tidal debris of disrupted
satellites whose orbital planes were roughly coincident with that of
the disk. This result prompted us to analyze the evolution of
satellite debris close to the galactic plane to explore whether it
could explain the ``ring-like'' structure discovered by the SDSS.

The formation of ripples, arcs, and rings are common phenomena
associated to mergers of galaxies. They arise from the folding and
wrapping of tidally-stripped stars in the potential well of the host
galaxy.  In this Letter we show that the structure observed by Y03 and
I03 could indeed be a tidal feature, and as such it may be considered
a natural consequence of the hierarchical build-up of galaxies.

\section{Tidal arcs in a cosmological simulation}
\label{sec:tarc}

\begin{figure*}[t]
\begin{center}
\includegraphics[height=0.75\textheight,clip,angle=90]{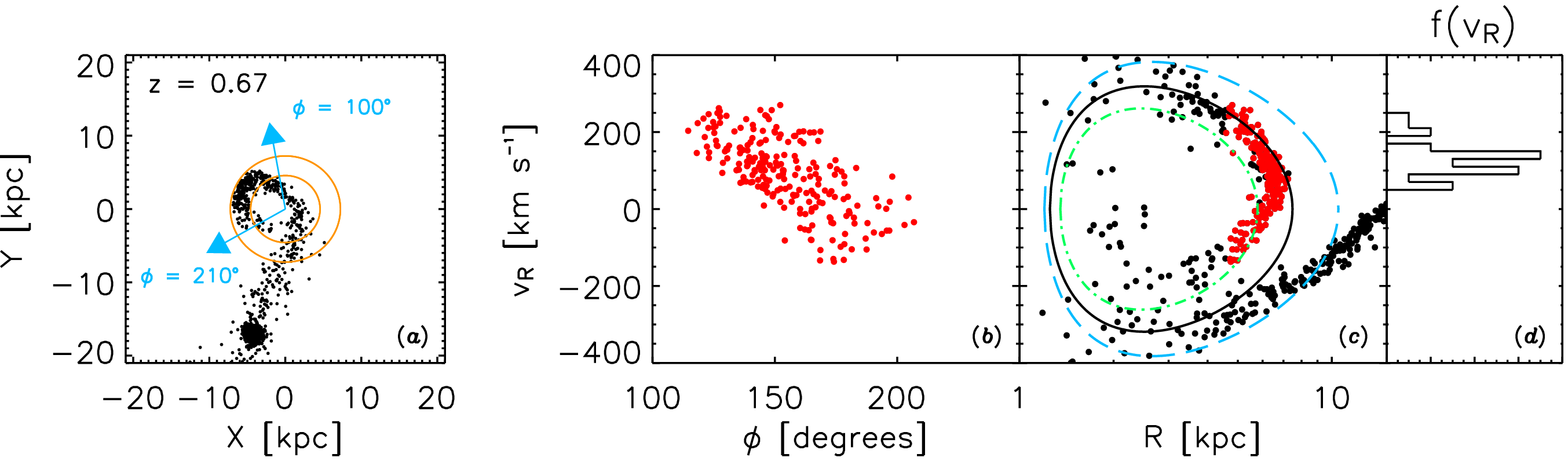}
\end{center}
\begin{center}
\includegraphics[height=0.75\textheight,clip,angle=90]{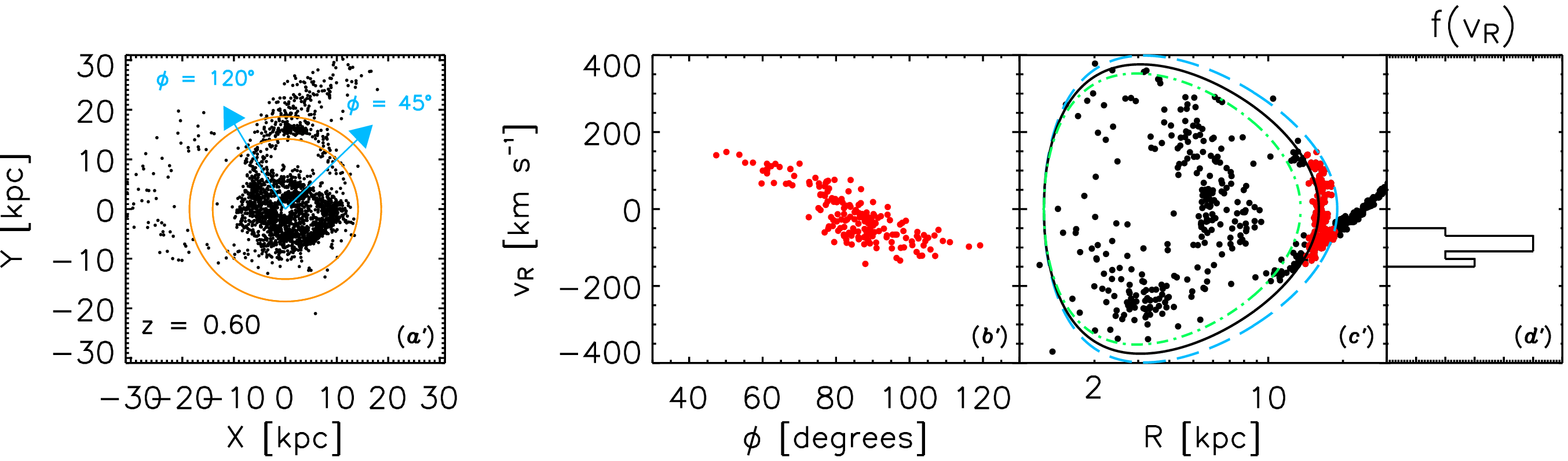}
\end{center}
\caption{Snapshots of the disruption of a satellite on an orbit
roughly coplanar to the disk, taken from the simulation of Abadi et
al. (2002a,b). Top and bottom rows correspond to two different times
in the simulation, $z=0.67$ and $z=0.60$, respectively. Panels ({\em
a}) and ({\em a}') show a ``face on'' projection of the debris. The
main galaxy is at the center of coordinates, but only the stars in the
satellite are shown for clarity. Panels ({\em b}) and ({\em b}') show,
in a cylindrical coordinate system, the (galactocentric) radial
velocity versus azimuthal angle $\phi$ for particles in the arc
delimited in radius by the two concentric circles and in longitude by
the arrows labeled by $\phi$. Panels ({\em c}) and ({\em c}') show
radial velocity and distance for all satellite particles (top row) or
for those with $45\arcdeg\! < \!\phi\! < \!120\arcdeg$ (bottom row),
respectively. Curves in these panels indicate the loci of particles
with three different values of the binding energy, $E$, and angular
momentum equal to the average of particles in the arc. Panels ({\em
d}) and ({\em d}') show the velocity distribution for particles in the
annuli at $ 130\arcdeg \le \phi \le 140\arcdeg$ and $100\arcdeg \le
\phi \le 110\arcdeg$, respectively.\label{fig:tarc}}
\end{figure*}

Figure~\ref{fig:tarc} illustrates the evolution of the debris of a
tidally disrupted satellite in the cosmological simulation of the
formation of a disk galaxy described by Abadi et al. (2002a,b).  As
discussed by these authors, this galaxy resembles more an early-type
spiral than the Milky Way; still, the luminosity of the disk component
($L_{\rm disk} \sim 2 \times 10^{10} \sL$) and the circular speed at
$R=10$ kpc ($\sim 250$ \kms) are comparable to those of our
Galaxy. Hence, physical quantities quoted here may be taken as
indicative of similar accretion events affecting the Galaxy without
substantial rescaling. We emphasize, however, that the simulated
galaxy is {\it not} a model of the Milky Way, and that our main goal
is to understand {\it qualitatively} rather than quantitatively the
nature of the structure observed by Y03 and I03.  Indeed, the
simulation precedes the announcement of the ring by Y03, and its
results were thus in no way prejudiced by its discovery.

The satellite has an orbit roughly coplanar with that of the
main disk of the galaxy, and its ``face on'' projection is shown
in the left panels of Figure~\ref{fig:tarc}.  The accretion of this
satellite adds $3.5 \times 10^9 M_\sun$ (or about $6\%$) to the
stellar mass of the preexisting galaxy. The first pericentric
passage that results in significant stripping occurs at $z\sim 0.67$.


As is clear from Figure~\ref{fig:tarc}, sharp ``ring-like'' structures
are readily seen at various times during the disruption of the
satellite.  In the top panels of Fig.~\ref{fig:tarc}, at $z=0.67$,
most stars with $4.6$ kpc $\! < \!R\! < \! 7.2$ kpc and $100\arcdeg\!
< \!\phi\! < \!210\arcdeg$ were stripped from the satellite during its
first pericentric passage.  The top panels ({\em b}) and ({\em c})
demonstrate that this incomplete ring (hereafter ``tidal arc'', for
short) is made out of particles with a range of energies that decrease
systematically with increasing angle $\phi$: particles at $\phi\sim
175\arcdeg$ are at the apocenter of their orbit, whilst those at
$\phi\sim 125\arcdeg$ have yet to reach it, and those at $\phi \sim
200\arcdeg$ are already past it. Their coincidence on an ``arc'' of
roughly constant radius is therefore a transitory event. Similar arcs
are produced after every pericentric passage that strips a significant
number of stars from the satellite; for example, a second one may be
seen at $z=0.60$ between $14.1$ kpc $\! < \!R\! < \! 18.6$ kpc, and
$45\arcdeg \le \phi \le 120\arcdeg$, as shown in panel ({\em a}') of
Fig.~\ref{fig:tarc}.  These ``tidal arcs'' become progressively less
well defined with time, as the orbits of particles in different energy
levels precess around the main galaxy at different rates, weakening
the radial coincidences shown in Fig.~\ref{fig:tarc}. After $z=0.5$,
tidal arcs become very difficult to discern in the satellite debris in
our simulation.

The tidal arc formation mechanism described above implies a number of
well defined properties that may be verified observationally. Perhaps
the most salient is illustrated in panels ({\em b}) and ({\em b}') of
Figure~\ref{fig:tarc}: {\it a well defined radial velocity gradient is
expected across the tidal arc}. For the two tidal arcs discussed thus
far, the (galactocentric) radial velocity varies by almost $200 $\kms
across $50$ degrees in azimuthal angle. At any given location, the
radial velocity distribution is fairly narrow, as shown in panels
({\em d}) and ({\em d}') of Fig.~\ref{fig:tarc}. Typical velocity
dispersions are of order $\sim 20$ to $50$\kms.

Another notable feature is that tidal arcs span a limited azimuthal
range. The actual range spanned depends both on the location of the
observer as well as on the eccentricity of the satellite's orbit at
the time of stripping; more circular orbits lead typically to wider
rings but somewhat less sharply defined in radius. Interestingly, the
angular extent of the tidal arcs shown in Figure~\ref{fig:tarc}
($\Delta\phi\sim 100\arcdeg$) is comparable to that of the structure
observed by Y03 and I03.  We also note that particles on the tidal
arcs are on rather eccentric orbits: the specific angular momentum of
particles in the tidal arc seen at $z=0.67$ is almost a factor of $2$
lower than that of a circular orbit at the same radial location, and
their $R_{\rm per}/R_{\rm apo}$ ratio averages $0.18$.

The fraction of the mass of the satellite contained within each arc
varies according to the fraction of mass stripped from the satellite
at each pericentric passage, which in turn depends on the orbit.
Since it is possible to strip a substantial fraction of the
satellite's mass in one single passage (see, e.g., Hayashi et
al. 2003), the rings might contain a significant fraction of the
original mass of the satellite: in particular, the arcs identified at
$z=0.67$ and at $z=0.60$ contain, respectively, $\sim 8\%$ and $\sim
7\%$ of the total mass of the satellite.

Finally, the orbit of the disrupting satellite is not exactly coplanar, leading
to asymmetries above and below the plane of the disk: the rings identified in
Figure~\ref{fig:tarc} at $z=0.67$ and $0.60$ have, respectively, $64\%$ and
$60\%$ of their mass above the plane of the disk, with a typical scaleheight of
$\sim 1$ kpc. In summary, tidal arcs are structures characterized by being
transitory, asymmetric in velocity and height, and of limited (but possibly
substantial) angular extent.

\section{Radial shells as relicts of minor mergers}
\label{sec:shell}

One potential difficulty for the tidal-arc interpretation of ring-like
features in the outer Galactic disk is that these arcs are transient
configurations that phase-mix away within a few rotation periods. The
actual timescale depends both on the initial properties as well as on
the orbital period $T$ of the satellite.  Typically the surface
density will decrease in time as $(T/t)^2$ (Helmi \& White 1999).  The
tidal arcs in Fig.~\ref{fig:tarc} have $T \sim 0.1$ Gyr, hence they
become a factor of ten fainter just one Gyr after the first
pericentric passage. This explains why they cannot be recognized
unequivocally in the cosmological simulation, particularly given our
limited numerical resolution.  We recall, however, that sharp features
and ripples in physical space are expected to be long-lived under most
circumstances, as shown, for example, by earlier work on the origin of
shells around elliptical galaxies (e.g. Quinn 1984; Dupraz \& Combes
1986; Hernquist \& Quinn 1988, 1989; Heisler \& White 1990).

To test the hypothesis that other ring-like structures might be
long-lived, we have followed the disruption of a $10^6$-particle
satellite (modeled after the one shown in Fig.~\ref{fig:tarc}) in a
fixed galactic potential. We use a quadrupole expansion code (Zaritsky
\& White 1988), and represent the satellite with a King profile of
concentration $\log_{10}(r_t/r_c) = 1.34$, total mass $3.44 \times
10^9 M_\sun$, total one-dimensional velocity dispersion $59.4$ km
s$^{-1}$, and core radius $r_c = 0.82$ kpc. The main galaxy is chosen
to match that of the cosmological simulation, and is approximated by
the superposition of a Miyamoto-Nagai disk (Miyamoto \& Nagai 1975), a
Hernquist bulge (Hernquist 1990) and an NFW halo (Navarro, Frenk \&
White 1997).  The initial position and velocity of the center of mass
of the satellite are those of the satellite in the cosmological
simulation at $z = 0.74$, before any significant disruption takes
place.


\begin{figure*}[t]
\centering\includegraphics[height=0.75\textheight,clip,angle=90]{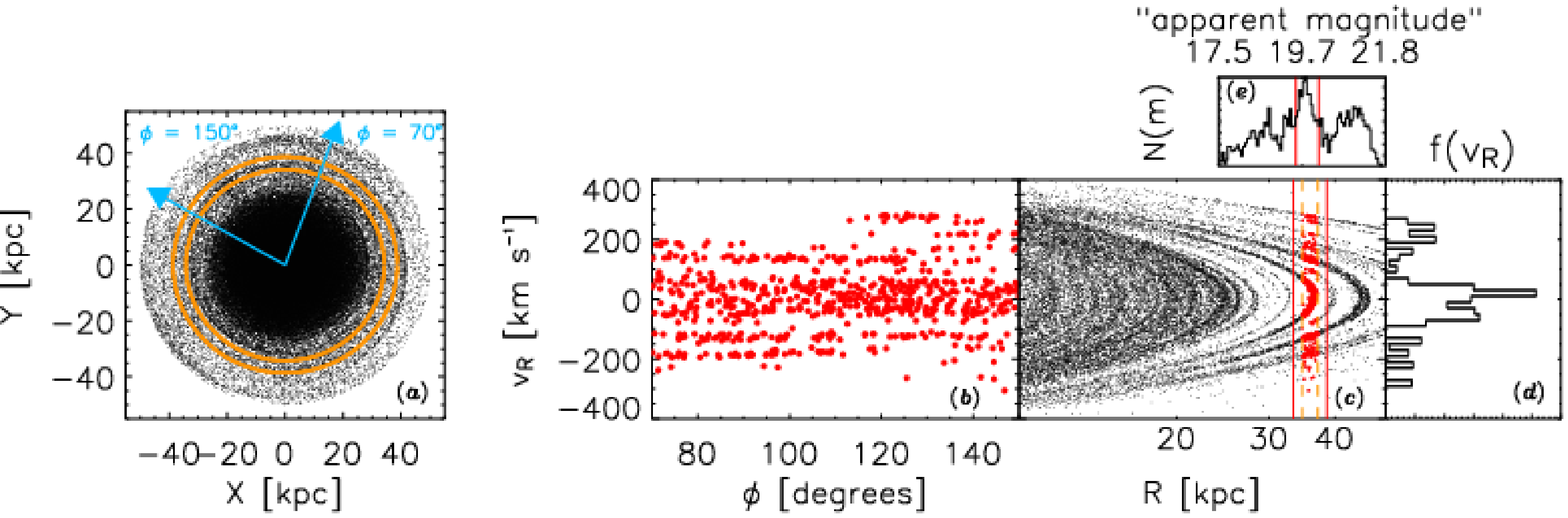}
\caption{Same as Figure~\ref{fig:tarc}, but for a simulation of the
disruption of a $10^6$-particle spherical satellite modeled after that
discussed in \S~\ref{sec:tarc}. The debris is shown at $z=0$ and
illustrates the persistence of density enhancements of constant radii
(``shells''). Panel ({\em b}) shows, using cylindrical coordinates,
the (galactocentric) radial velocity versus azimuthal angle $\phi$ for
particles in the annulus highlighted in panel ({\em a}). The vertical
dashed lines in panel ({\em c}) correspond to the radii defining the
annulus shown in the X-Y projection, while the solid lines indicate
the range of distances when a spread in magnitudes is assumed, as
shown in panel ({\em e}). Panel ({\em d}) shows the radial velocity
distribution for particles with ``apparent magnitudes'' in the range
$19.5- 20.1$ and $140\arcdeg \le \phi \le 150\arcdeg$.
\label{fig:shell}}
\end{figure*} 


Since the goal of this (re)simulation is to study the long-term
survival of ring-like features, we focus our analysis on the structure
of the debris at $z=0$ (7 Gyr after infall). Panel ({\em a}) in
Fig.~\ref{fig:shell} shows a ``face-on'' view of the satellite debris
at this time: several sharply-defined ring-like features are easily
identified in this panel, especially in the outer regions. To
illustrate the properties of these features, we choose the annulus
defined by $34.7$ kpc $ \! < \! \! R \!<\!  37.1$ kpc and $70\arcdeg\!
< \!\phi\!  < \!150\arcdeg$. The radial enhancement is dominated by
particles at the apocenter of their orbits (i.e., those with $v_R
\approx 0$ in panel ({\em c}) of Fig.~\ref{fig:shell}). This density
structure is a caustic surface (2-D singularity or fold) in the
mapping of the phase-space density of a cold/small system into the 3-D
physical space (Nulsen 1989; Hernquist \& Quinn 1988).  This structure
is analogous the ``shell'' structures found around some elliptical and
a few spiral galaxies (Malin \& Carter 1983; Schweizer \& Seitzer
1988; Wilkinson et al. 2000).

To assess the visibility of shells when a spread of magnitudes is
present, we have convolved the distance distribution of particles with
$140\arcdeg \le \phi \le 150\arcdeg$ and $R \ge 25$ kpc with the
magnitude distribution near the turnoff of the globular cluster M92,
i.e. for $0.16 \le (g-r) \le 0.3$ (Stetson \& Harris
1988)\footnote{The $B$ and $V$ magnitudes have been transformed to the
$g$ and $r$ passbands following Cohen (1985).}.  Here we assumed the
observer is located at $R \sim 25$ kpc, which implies that its
distance to the shell is the same as that of the structure observed by
SDSS. The apparent magnitude distribution $N(m)$ of these particles,
plotted in panel (2{\em e}), is similar to that shown in Newberg et
al. (2002).

Panel ({\em b}) of Figure~\ref{fig:shell} shows that there are no
gradients in the radial velocity across the ``shell''. The radial
velocity distribution at any given location on the shell has several
components, as can be readily seen from panels (2{\em c}) and (2{\em
d}): it is dominated by a central core with $|v_R| \lsim 100$\kms and
$\sigma_R\sim 40 $\kms comprising $\sim 70\%$ of the stars, together
with a number of small peaks corresponding to particles of higher
energy on their way to or from their (more distant) apocenters.  The
presence of these peaks at high $|v_R|$ increases the radial velocity
dispersion of the ring to $\sim 120 $\kms.

Shells are relaxed features that are expected to be symmetric above/below the
plane of the disk: roughly $50\%$ of the particles in the shell identified in
Fig.~\ref{fig:shell} lie above the disk plane. Shells are long-lived, but it is
unlikely that they may carry a significant fraction of the original mass of the
satellite: the amount of mass they contain decreases rapidly as $(T/t)^3$. The
radial shell shown in Fig.~\ref{fig:shell} contains only $0.7\%$ of the
satellite's mass, an order of magnitude smaller than the fraction of mass
carried by the tidal arcs shown in Fig.~\ref{fig:tarc}. Other long-lived shell
structures that are easily identified in the same figure carry a similarly low
fraction ($\lsim 1\%$) of the original mass of the satellite. 

\section{Discussion}
\label{sec:disc}

The main purpose of the simulations presented here is not to match
quantitatively the properties of the structure discovered by SDSS, but
rather to show that qualitatively similar features arise naturally as
a result of the accretion of satellites on orbits roughly coplanar
with the Galactic disk. Tidal arcs (\S~\ref{sec:tarc}) are prevalent
during the disruption process (which lasts a few orbital times, or
$\lesssim 1$ Gyr) whereas shells (\S~\ref{sec:shell}) are long-lasting
features that may be identified long after the debris has reached
equilibrium in the main galactic potential.

A cursory comparison with the properties of the structure observed by
Y03 and I03 shows that both interpretations are consistent with
observations. Indeed, the extent in Galactic longitude is comparable
to both arcs and shells shown in Figs.~\ref{fig:tarc} and
\ref{fig:shell}. The systematic variation in the mean radial velocity
with longitude claimed by Y03, from $22$ km s$^{-1}$ at
$(l,b)=(182\arcdeg, 28\arcdeg)$, to $103$ km s$^{-1}$ at $(l,b) =
(225\arcdeg, 27\arcdeg)$, could be, but is not necessarily, a
signature of a ``tidal arc''. This ``gradient'' can also be entirely
attributed to perspective effects resulting from the projection of the
solar motion and the tangential velocity of stars in the ring along
these two different lines of sight. This degeneracy can be broken with
radial velocity measurements along more widely separated lines of
sight, and a gradient, if detected, would then favor the tidal arc
interpretation. On the other hand, larger spectroscopic samples would
allow to test whether the radial velocity distribution is smooth or if
it consists of multiple peaks as in the shell scenario.  Measurements
of the extent of the structure in Galactocentric longitude, and
above/below the plane would provide other possible ways of
distinguishing amongst the two interpretations.

Discriminating between these two scenarios is important in order to
estimate robustly the time of accretion as well as the original mass
of the satellite.  The tidal arc interpretation would suggest a very
recent accretion event of a fairly small satellite; if the arc
comprises $\sim 10\%$ of the original mass of the satellite, the mass
estimated for the ring by Y03 would imply a progenitor mass of roughly
$2.5 \times 10^8 \sm$.  Thus, the Sagittarius dwarf galaxy (Ibata,
Gilmore \& Irwin 1994) might not be the only satellite being accreted
today by the Milky Way.  The shell scenario, on the other hand, would
imply the past accretion of a satellite roughly ten times more massive
($2.5 \times 10^9 \sm$), and the presence of considerable debris in
the inner (thick) disk. Such stars are expected to have quite distinct
kinematics and to be on rather eccentric orbits, perhaps consistent
with the peculiar kinematics of the thick disk in the solar
neighborhood reported by Gilmore, Wyse \& Norris (2002).

We end by noting that tidal ripples of the kind discussed here are not
a new idea, and were discussed by Hernquist \& Quinn (1988) and others
well before our study. The cosmological simulation analyzed here
suggests that satellites will often have orbits approximately coplanar
with the disk of the primary, and thus a significant fraction of their
debris should be concentrated in the Galactic disks (Abadi et
al. 2002b).  Identifying and cataloging such structures will be
possible with large surveys like RAVE (Steinmetz 2002) and GAIA
(Perryman et al. 2001), and should help us unravel the puzzle of the
assembly of the Galaxy.

\acknowledgments This work has been partially supported by grants from
the U.S. National Aeronautics and Space Administration (NAG 5-10827)
and the David and Lucile Packard Foundation.

\end{document}